\documentclass[sigconf]{acmart}%

\usepackage{pifont}
\usepackage{subfig}
\usepackage{multirow}
\AtBeginDocument{%
  \providecommand\BibTeX{{%
    \normalfont B\kern-0.5em{\scshape i\kern-0.25em b}\kern-0.8em\TeX}}}

\begin{document}

\newcommand{\nt}[1]{{\sout{#1}}}

\newcommand{\st}[1]{}

\newcommand{\clst}{{\textit{CL List}}}

\newcommand{\alst}{{\textit{Dependence List}}}

\newcommand{\nwtxt}[1]{{\textcolor{black}{#1}}}
\newcommand{\newtext}[1]{{\textcolor{blue}{#1}}}
\newcommand{\chkTxt}[1]{{\textcolor{blue}{#1}}}
\newcommand{\pmem}{{persistent memory}}
\newcommand{\sota}{{HWUndo}}
\newcommand{\sotard}{{HWRedo}}
\newcommand{\lp}{{LPO}}
\newcommand{\dtp}{{DPO}}
\newcommand{\Eg}{{\textit{E.g.}}, }
\newcommand{\eg}{{\textit{e.g.}}, }
\newcommand{\ie}{{\textit{i.e.}}, }
\newcommand{\etal}{{\textit{et al.}}}
\newcommand{\arch}{{\textit{ASAP}}}
\newcommand{\archTitle}{ASAP}
\newcommand{\archnobold}[1]{{{{\textit{ASAP}}}}{#1}}
\newcommand{\archabv}{{{hula_pm}}}
\newcommand{\gref}[1]{\textbf{G\ref{#1}}}

\newcommand{\aref}[1]{\textbf{A\ref{#1}}}
\newcommand{\contref}[1]{\textbf{C\ref{#1}}}

\newcommand{\website}[1]{{\tt #1}}
\newcommand{\program}[1]{{\tt #1}}
\newcommand{\benchmark}[1]{{\it #1}}
\newcommand{\fixme}[1]{{\textcolor{red}{\textit{#1}}}}

\newcommand*\circled[2]{\tikz[baseline=(char.base)]{
            \node[shape=circle,fill=black,inner sep=1pt] (char) {\textcolor{#1}{{\footnotesize #2}}};}}

\ifx\figurename\undefined \def\figurename{Figure}\fi
\renewcommand{\figurename}{Figure}
\renewcommand{\paragraph}[1]{\textbf{#1}~~}
\newcommand{\figline}{{\vspace*{.05in}\hline}}

\newcommand{\Sect}[1]{$\S$~\ref{#1}}
\newcommand{\Fig}[1]{Figure~\ref{#1}}
\newcommand{\Tbl}[1]{Table~\ref{#1}}
\newcommand{\Equ}[1]{Equation~\ref{#1}}
\newcommand{\Apx}[1]{Appendix~\ref{#1}}

\newcommand{\specialcell}[2][c]{\begin{tabular}[#1]{@{}l@{}}#2\end{tabular}}

\newcommand{\RNum}[1]{\uppercase\expandafter{\romannumeral #1\relax}}


\newcommand{\ignore}[1]{}

\iffalse
\newcommand{\todo}[1]{{\color{red}[\textbf{\sc TODO}: \textit{#1}]}}
\newcommand{\kim}[1]{{\color{red}[\textbf{\sc nskim}: \textit{#1}]}}
\newcommand{\izzat}[1]{{\color{green}[\textbf{\sc Izzat}: \textit{#1}]}}
\newcommand{\ahmed}[1]{{\color{blue}[\textbf{\sc Ahmed}: \textit{#1}]}}
\newcommand{\kimNw}[1]{{\color{red}{#1}}}
\newcommand{\izzatNw}[1]{{\color{green}{#1}}}
\newcommand{\ahmedNw}[1]{{\color{blue}{#1}}}
\newcommand{\addition}[1]{{\color{purple}#1}}
\newcommand{\rmtxt}[1]{{\textcolor{red}{#1}}}
\else
\newcommand{\todo}[1]{}
\newcommand{\nskim}[1]{}
\newcommand{\izzat}[1]{}
\newcommand{\ahmed}[1]{}
\newcommand{\izzatNw}[1]{#1}
\newcommand{\addition}[1]{#1}
\newcommand{\rmtxt}[1]{}
\fi

\newcommand{\newCont}[1]{{\textcolor{red}{#1}}}

\title{Asynchronous Persistence with \archTitle}
\titlenote{This work has been published at ISCA '22~\cite{ASAP}.}
\author{Ahmed Abulila}
\affiliation{%
  \institution{Microsoft Corporation}
  \streetaddress{}
  \city{}
  \country{}
}

\author{Izzat El Hajj}
\affiliation{%
  \institution{American University of Beirut}
  \streetaddress{}
  \city{}
  \country{}}

\author{Myoungsoo Jung}
\affiliation{%
  \institution{Korea Advanced Institute of Science and Technology}
  \city{}
  \country{}
}

\author{Nam Sung Kim}
\affiliation{%
 \institution{University of Illinois at Urbana-Champaign}
 \streetaddress{}
 \city{}
 \state{}
 \country{}}





\maketitle
\section{Motivation}\label{sec:motivation}

\fbox{\parbox{0.97\columnwidth}{\centering
The problem our work tackles is hardware logging for \pmem{}. It is important because it promises higher performance than pure software solutions.
}}\vspace{0.1cm}

Persistent memory provides the byte-addressability and low latency of DRAM as well as the persistency of storage devices. 
Programming persistent data structures typically involves writing atomic regions that are atomic and durable, ensuring that stores to persistent memory within an atomic region persist in an all-or-none manner.
To guarantee the atomicity and durability of atomic regions, Write-Ahead Logging (WAL) has commonly been used~\cite{lognvmm}.
WAL consists of two key persist operations: log persist (\lp) and data persist (\dtp) \cite{aries}.
\lp{} operations flush log entries to \pmem{} before making the data persistent, to ensure that a consistent state can be recovered if a crash occurs before all the data of a given atomic region has persisted.
\dtp{} operations write back the actual data modified by the atomic region to \pmem{}.

Software logging solutions~\cite{mnemosyne} for \pmem{} offload the complexity of managing logs to the software, placing persist operations on the critical path of an atomic region's execution.
In contrast, hardware logging solutions~\cite{ReDU, proteus} for \pmem{} can initiate persist operations in a manner that is transparent to software, and can complete these operations in the background, overlapping them with the execution of other instructions and ultimately resulting in better performance.


\vspace{-2ex}

\section{Limitations of the State of the Art}\label{sec:limitations}

\fbox{\parbox{0.97\columnwidth}{\centering
The key limitation of state-of-the-art hardware solutions is that hardware \textit{undo} logging solutions do not support asynchronous persist operations, whereas hardware \textit{redo} logging solutions support asynchronous data persist operations but not asynchronous log persist operations.
}}\vspace{0.1cm}

Persist operations that are overlapped with execution of other instructions can be classified as \textit{synchronous} or \textit{asynchronous} with respect to the end of an atomic region.
Synchronous persist operations are overlapped with execution of other instructions \textit{within} an atomic region, but once the end of an atomic region is reached, all persist operations initiated by the atomic region must complete before instruction execution proceeds.
In contrast, asynchronous persist operations are also overlapped with instructions that are executed \textit{after} the atomic region, hence allowing instruction execution to proceed past the end of an atomic region without waiting for persist operations to complete.

The advantage of asynchronous persist operations is that they reduce idle time by not waiting at the end of an atomic region.
Moreover, in a multi-threaded context, if the atomic region is nested inside a critical section to guarantee isolation,
the latency of asynchronous persist operations will not be part of the critical section~\cite{PMEMSpec}.
Removing the persistence latency from critical sections benefits concurrency by reducing the execution time of critical sections.

The two major approaches to hardware WAL are undo-logging~\cite{proteus} and redo-logging~\cite{ReDU}.
While undo logging has several advantages over redo logging which are discussed in the paper, a key disadvantage is that hardware solutions for undo logging do not support asynchronous \dtp{} operations, whereas hardware solutions for redo logging do.
Neither support asynchronous \lp{} operations.

\vspace{-2ex}


\section{Key Insights}\label{sec:key-insights}

\fbox{\parbox{0.97\columnwidth}{\centering
The key insight in the paper is that asynchronous log and data persist operations can both be supported under undo logging if control and data dependences between atomic regions are tracked and enforced.
}}\vspace{0.1cm}

The challenge with supporting asynchronous persist operations under undo logging is that allowing instruction execution to proceed past the end of an atomic region before the atomic region has completed its persist operations may result in violating control and data dependencies between atomic regions.
In other words, it runs the risk of a later atomic region committing before an earlier one does, or, in a multi-threaded context, a consumer atomic region committing before the corresponding producer atomic region does~\cite{DATM}.
These situations leave the data in an unrecoverable state after a crash.
The paper includes examples that demonstrate these situations.

We show that asynchronous \lp{} and asynchronous \dtp{} operations can both be supported under undo logging by tracking control and data dependences between atomic regions and enforcing atomic regions commit (undo logs are freed) in a manner that respects these dependences.

\vspace{-2ex}

\section{Main Artifacts}\label{sec:main-artifacts}

\fbox{\parbox{0.97\columnwidth}{\centering
The main artifact presented in the paper is a hardware design that supports undo logging with asynchronous log and data persist operations by tracking and enforcing control and data dependencies between atomic regions in hardware.
}}\vspace{0.1cm}

\begin{figure}[t]
    \centering
    \includegraphics[width=0.96\columnwidth]{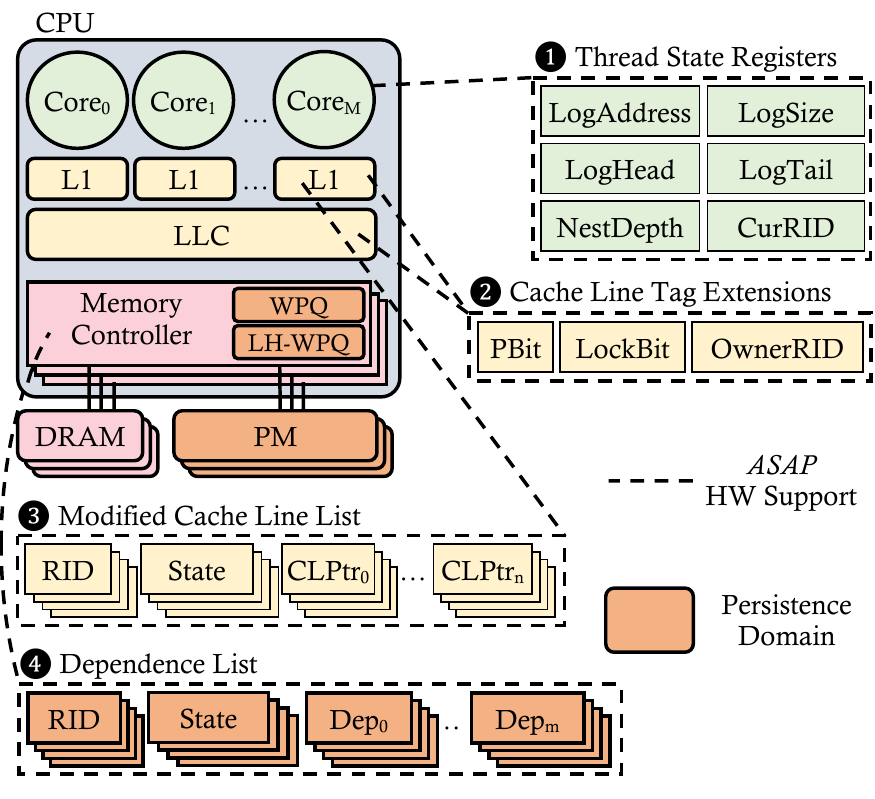}
    \vspace{-2ex}
    \caption{\arch{} hardware extensions}
    \vspace{-4ex}
    \label{fig:hw-data-structures}
\end{figure}

The paper presents \arch{}, the first \textbf{\underline{A}}rchitecture \textbf{\underline{S}}upport for \textbf{\underline{A}}synchronous \textbf{\underline{P}}ersistence (\arch{}).
\arch{} allows both \lp{} and \dtp{} operations to happen asynchronously, and tracks control and data dependencies between atomic regions in hardware to ensure that atomic regions commit (undo logs are freed) in the proper order.
The hardware extensions that \arch{} requires are shown in \Fig{fig:hw-data-structures}.
The key extensions include:
\begin{itemize}
    \item Per-thread registers (\ding{182}) assist with 
log management
    \item Structures to track which atomic region last modified a persistent cache line to help capture data dependecies between atomic regions (\ding{183})
    \item Structures to track which cache lines have been modified by an atomic region to ensure that all the cache lines persist before the atomic region commits (\ding{184})
    \item Structures to track which atomic regions are still active and the atomic regions they depend on to ensure that all an atomic region's dependencies have been resolved before its log is freed (\ding{185})
\end{itemize}
\arch{} provides a simpler software interface than prior hardware undo logging approaches.
Also, \arch{} applies three key optimizations to reduce persistent memory traffic: \lp{} dropping, \dtp{} coalescing, and \dtp{} dropping.
These optimizations are particularly effective in combination with asynchronous persist operations in reducing memory traffic.
\arch{} is implemented and evaluated using gem5.

\vspace{-1ex}

\section{Key Results and Contributions}\label{sec:key-contributions}

\fbox{\parbox{0.97\columnwidth}{\centering
We show that \arch{} improves performance compared to state-of-the-art hardware logging approaches that perform some persist operations synchronously.
}}\vspace{0.1cm}

\ignore{
\textcolor{blue}{
\begin{itemize}
  \item What are the most important \emph{one or two} empirical or theoretical
    results of this approach?
  \item What are the contributions that this paper makes to the state of the
    art? List them in an \texttt{itemize} section. Each contribution should be no more than a few sentences long.
  \item Clearly describe its advantages over past work, including how it overcomes their limitations.
\end{itemize}
}
}

\begin{figure}[t]
    \centering
        \includegraphics[width=\columnwidth]{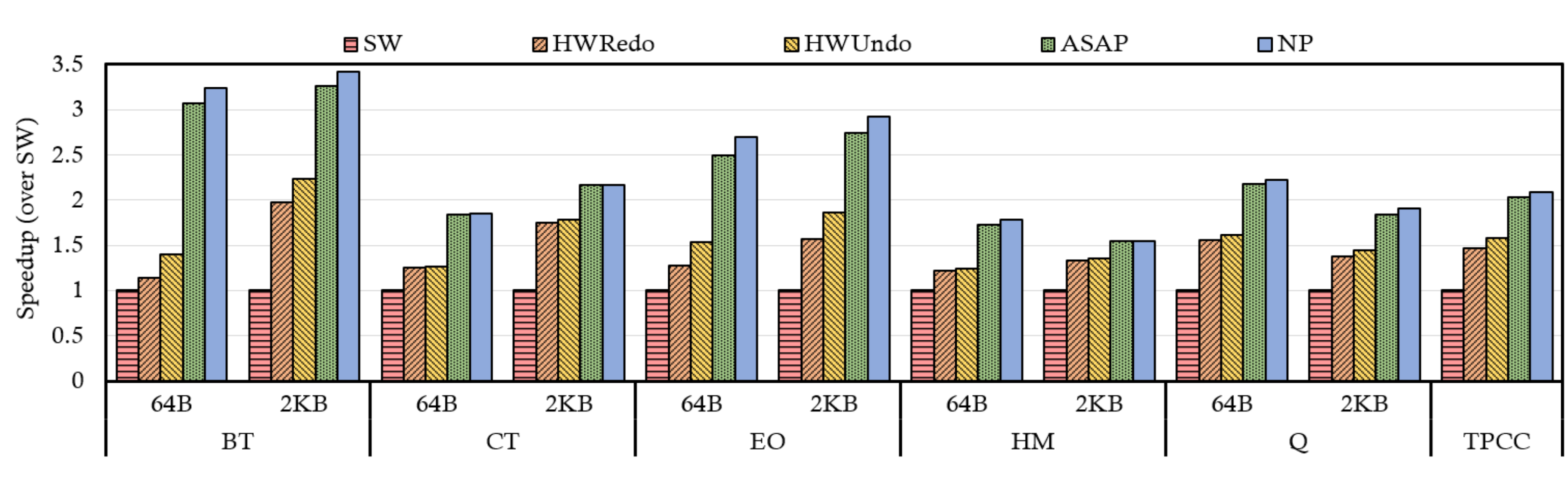}
        \vspace{-4ex}
        \caption{Performance comparison (speedup over SW) for select benchmarks}
        \vspace{-4ex}
    \label{fig:eval-time}
\end{figure}

We compare \arch{} to the following baselines:
\begin{itemize}
    \item \textbf{SW:} Software-only implementation of undo-logging
    \item \textbf{\sota{}:} Based on the state-of-the-art hardware undo logging implementation~\cite{proteus} which performs \lp{} and \dtp{} operations synchronously
    \item \textbf{\sotard{}:} Based on the state-of-the-art hardware redo logging implementation~\cite{ReDU} which performs \lp{} operations synchronously
    \item \textbf{NP:} Data is read from and written to persistent memory, but no crash consistency is guaranteed
\end{itemize}
The performance improvement for select benchmarks is shown in Figure~\ref{fig:eval-time}.
More benchmarks are evaluated in the paper.
\arch{} improves performance compared to state-of-the-art hardware undo logging by 1.41$\times$ (geomean), while achieving 0.96$\times$ (geomean) of the ideal performance when no persistence is enforced.
\arch{} also reduces persistent memory traffic, generating 0.52$\times$ (geomean) the traffic generated by state-of-the-art hardware undo logging techniques.
\arch{} is robust against increasing persistent memory latency, and is therefore suitable for both fast and slow persistent memory technologies.

\vspace{-1ex}

\section{Conclusions}\label{sec:conclusions}

This paper presents \arch{}, a hardware logging scheme that allows atomic regions to commit asynchronously.
Committing atomic regions asynchronously removes the need to wait for log persist and/or data persist operations at the end of atomic regions, which reduces the latency of these regions.
To ensure that the atomic regions commit in the proper order, \arch{} tracks and enforces control and data dependencies between atomic regions in hardware.
Our evaluation shows that \arch{} outperforms state-of-the-art hardware undo and redo logging techniques, which commit atomic regions synchronously.
It also reduces persistent memory traffic and is suitable for both fast and slow persistent memory technologies.

\vspace{-1ex}

\balance
\bibliographystyle{IEEEtranS}
\bibliography{references}
\end{document}